\pdfoutput=1
\documentclass[preprint,12pt]{elsarticle}

\usepackage[utf8]{inputenc}
\usepackage{amssymb}
\usepackage{algorithmic}
\usepackage{algorithm}

\begin{document}

\begin{frontmatter}

\title{An Exact Parallel Dense Gauss-Seidel Algorithm with Trade-Offs between Parallel Execution Time and Preprocessing Time\tnoteref{t1}}

\author[upb]{Mugurel-Ionu\c t Andreica}
\author[upb]{Pantelimon George Popescu}
\author[upb]{Eliana-Dina T\^ ir\c sa}

\address[upb]{Computer Science Department\\
Politehnica University of Bucharest\\
Splaiul Independen\c tei 313, sector 6, 060042, Bucharest, Romania}

\begin{abstract}
The Gauss-Seidel iterative method is commonly used for solving systems of linear equations because of its good convergence properties. However, the Gauss-Seidel algorithm cannot be easily parallelized because of the data dependencies existing at each iteration. The best parallel execution time for dense systems of equations is $O(n)$ given that enough processors are available and no preprocessing is performed. In this paper we propose a new parallel Gauss-Seidel algorithm which is capable of achieving better parallel runtimes at the expense of more preprocessing time. The algorithm can be tuned in order to provide convenient trade-offs between the preprocessing time and the best parallel runtime achievable by the algorithm.
\end{abstract}

\tnotetext[t1]{This manuscript was originally written in July 2013, and the names and affiliation of the authors are reflecting the state at that time. Recently, the original manuscript went through a few rounds of AI-based reviews, which identified several typos and bugs, which were corrected, but which didn't change the original ideas. AI was only used for identifying issues, not for correcting them.}

\end{frontmatter}

\section{Introduction}
\label{intro}

In order to solve large systems of linear equations, iterative methods like Jacobi or Gauss-Seidel \cite{jgs} are used. These methods produce good approximate results much faster than solving the system of equations exactly. When implemented sequentially, the Gauss-Seidel method is known to converge faster than the Jacobi method and, thus, it is preferable to use it. However, when the size of the system of equations is large enough even sequential implementations of these methods are too slow and parallel versions are required. The Jacobi method can be easily parallelized, because the value of each unknown at the current iteration depends only on the values of the unknowns at the previous iteration. The Gauss-Seidel method, on the other hand, seems difficult to parallelize because the value of an unknown $i$ at iteration $k$, $x(k,i)$, depends on all the values $x(k,j<i)$. This property is what helps the Gauss-Seidel method converge to a good solution with a smaller number of iterations than Jacobi, but it is also what prevents it from being easily parallelized. In this paper we propose a new parallel Gauss-Seidel algorithm for dense systems of linear equations which can overcome the limitations of the standard Gauss-Seidel method at the expense of more preprocessing time. In fact, with enough preprocessing time available and enough available processors, we can reach a parallel running time of $O(\log(n))$ per iteration ($n$ is the size of the system of linear equations), which is also the best result possible for the parallel version of the Jacobi method. However, this is not the main contribution of this paper. The most important result is that our algorithm can make use of a parameter $m$ which can tune the amount of preprocessing time and the amount of parallelism available at each iteration of the Gauss-Seidel method. The algorithm will provide a trade-off between the preprocessing time and the best achievable parallel runtime: the more preprocessing time we can spend, the more parallelism per iteration we can achieve (and, thus, the better the parallel runtime can be).

The rest of this paper is structured as follows. In Section \ref{gsmet} we describe the Gauss-Seidel method in detail and we discuss some parallelization opportunities of the standard formulation of this method. In Section \ref{pargsalgo} we present our new parallel Gauss-Seidel algorithm together with a detailed complexity analysis. The algorithm is focused on achieving the best parallel running time possible for a given number $P$ of available processors. In Section \ref{conclusions} we conclude.

\section{The Gauss-Seidel Method}
\label{gsmet}

Gauss-Seidel is an iterative method for solving systems of linear equations

\begin{equation}\label{eq1}
A\cdot x = b.
\end{equation}

$A$ is an $n\cdot n$ matrix, $x$ and $b$ are vectors with $n$ elements, and $x$ is the vector of unknowns. The method starts with an initial solution vector and after each iteration computes a new solution vector. The algorithm ends when the difference between the solution vector at the current iteration and the one from the previous iteration is below a predefined threshold. We define by $x(0)$ the initial solution vector and by $x(k)$ the solution vector obtained after $k$ iterations. If we write $A$ as $A=D-L-U$, where $D$ is the diagonal part of $A$, $-L$ is the strict lower part of $A$, and $-U$ is the strict upper part of $A$, then we can write $x(k+1)$ as

\begin{equation}\label{eq2}
x(k+1) = (D-L)^{-1} \cdot U \cdot x(k) + (D-L)^{-1} \cdot b
\end{equation}

If we denote by $A(i,j)$ the element at row $i$ and column $j$ of $A$, by $x(k,i)$ the element $i$ of $x(k)$ and by $b(i)$ the element $i$ of $b$ ($0\leq i, j\leq n-1$), then we can compute $x(k+1)$ as:

\begin{equation}\label{eq3}
x(k+1,i) = (b(i) - \sum_{j<i}{A(i,j)\cdot x(k+1, j)} - \sum_{j>i}{A(i,j)\cdot x(k, j)}) / A(i,i)
\end{equation}

The Gauss-Seidel method seems an inherently sequential method. $x(k,i)$ can be computed only after all the values $x(k,j<i)$ were already computed. The only obvious part which can be parallelized is the computation of each $x(k,i)$ value (which consists of $O(n)$ arithmetic operations). This computation (a reduction operation applied to $n$ values) can make use of $P\leq n$ processors and can be implemented with a time complexity of $O(n/P+\log(P))$ \cite{parred}. Thus, a Gauss-Seidel iteration can be very simply parallelized to run in $O(n^2/P+n\cdot \log(P))$ time but can use only at most $n$ processors. For $P=n$ the parallel runtime is $O(n\cdot \log(n))$. Note that in this paper we will use the terms and notations from \cite{parbook} regarding the analysis of parallel algorithms (in particular, the terms {\bf parallel runtime} and {\bf speedup}).

\section{The Parallel Gauss-Seidel Algorithm}
\label{pargsalgo}

In \cite{gsdense} the authors were able to devise other more intelligent parallelization schemes than the obvious parallel algorithm mentioned in the previous section. These algorithms are suitable in GPU settings (where a large number of processors is available). However, none of the parallel algorithms presented in \cite{gsdense} has a parallel runtime better than $O(n)$ per iteration no matter how many processors are available. In this section we present a parallel Gauss-Seidel algorithm which is capable of achieving a sub-linear time complexity per iteration. However, we are capable of obtaining this time complexity at the expense of more preprocessing time. Note that obtaining the maximum degree of parallelism is not conceptually difficult. All we need to do is to compute $E=(D-L)^{-1}\cdot U$ and $F=(D-L)^{-1}\cdot b$ and then use Eq. \ref{eq2}. In Eq. \ref{eq2} each component of $x(k+1)$ can be computed independently based only on the components of $x(k)$. Thus, for $P\leq n$ processors we can obtain a parallel runtime per iteration of $O(n^2/P)$. For $n<P\leq n^2$ the achievable parallel runtime per iteration is $O(n^2/P+\log(P/n))$.

In order to obtain the parallel runtimes mentioned we need to first compute the inverse of the lower triangular matrix $(D-L)$ and then multiply it by the upper triangular matrix $U$. Algorithms with $O(n^3)$ serial execution times exist for both operations. Parallel versions of these algorithms also exist. In \cite{partrimatinv} an algorithm with a parallel runtime of $O(n^3/P)$ for the inversion of triangular matrices has been proposed. However, the algorithm is only capable of using at most $n$ processors for inverting an $n\cdot n$ matrix. An $O(n)$ parallel runtime algorithm with $O(n^2)$ processors for inverting a triangular $n\cdot n$ matrix was obtained in \cite{parmatinvgauss}. The best parallel runtime that can be achieved by a known algorithm for inverting a triangular $n\cdot n$ matrix is $O(\log^2(n))$ using $O(n^3)$ processors \cite{parmatinv}, \cite{parmatinv2}.

The situation is better for parallel matrix multiplication algorithms \cite{parmatmult}. There are several algorithms which can achieve a parallel runtime of $O(n^3/P)$ when the number of available processors $P$ is at most $n^2$. An algorithm which can use $n^3$ processors and achieve an $O(\log(n))$ parallel running time was also analyzed in \cite{parmatmult} and improved in order to achieve this running time with only $n^3/\log(n)$ processors. Actually, the mentioned algorithm has a parallel runtime of $O(n^3/P+\log(P/(n^2)))$ when $n^2\leq P\leq n^3$ processors are used.

In general these computations may be prohibitive in terms of running time (in order to achieve very good parallel runtimes a large number of processors is required, which may not be available). The main contribution of this paper is to present a parallel Gauss-Seidel algorithm which can be tuned through a parameter $m$ ($1\leq m\leq n$). $m$ controls the amount of parallelism available per iteration and the amount of preprocessing time - larger values of $m$ lead to a larger degree of parallelism at the expense of more preprocessing time. In fact, $m=1$ will denote a rather efficient Gauss-Seidel method with $O(n)$ parallel runtime per iteration at best and $m=n$ will represent the algorithm described earlier with the maximum degree of parallelism ($O(\log(n))$ parallel runtime per iteration at best).

We will split the $n$ unknowns into groups of size $m$. We will assume that $m$ is a divisor of $n$. If not then we can increase $n$ to the nearest multiple of $m$. We will also add extra rows and columns to the matrix $A$ (all the elements on these extra rows and columns will be $0$, except on the main diagonal where we will set the elements equal to $1$) and extra elements to the vectors $b$ and $x$. The original unknowns will not depend on the extra unknowns (whose values we are not interested in). We will denote by $xg(k,i)$ the unknowns of group $i$ after $k$ iterations ($xg(k,i)$ consists of the values $x(k,m\cdot i),\ldots,x(k,m\cdot(i+1)-1)$). We define similarly $bg(i)$ as the $i^{th}$ group of $m$ consecutive values from the vector $b$.

In the preprocessing stage we will compute an $n/m$ by $n/m$ matrix $T$, where each element is a block of size $m\cdot m$. We will denote by $T(i,j)$ the block at row $i$, column $j$. We will also introduce the vectors $bgt(i)$, corresponding to the $bg(i)$ vectors. With these notations we can rewrite Eq. \ref{eq3} at block-level:

\begin{equation}\label{eq4}
xg(k+1,i) = bgt(i) + \sum_{j<i}{T(i,j)\cdot xg(k+1, j)} + \sum_{j\geq i}{T(i,j)\cdot xg(k, j)}
\end{equation}

Basically, $T(i, j)$ encodes the dependencies of $xg(k + 1, i)$ on: (a) $xg(k + 1, j)$ for $j<i$, or (b) $xg(k, j)$ for $j\geq i$.

We will discuss the preprocessing stage and the implementation of the parallel Gauss-Seidel algorithm in two separate subsections.

\subsection{The Preprocessing Stage}
An $O(n^2\cdot m)$ algorithm for computing all the entries in the matrix $T$ can be designed as follows. For each group $i (0\leq i\leq n/m-1)$, we will compute all the entries $T(i,j) (0\leq j\leq n/m-1)$. First, we have $T(i,i)(0,0)=0$ and $T(i,i)(0,1\leq v\leq m-1)=-A(i\cdot m, i\cdot m + v) / A(i\cdot m, i\cdot m)$. Similarly, for $j\neq i$, we have $T(i,j)(0,0\leq v\leq m-1)=-A(i\cdot m, j\cdot m + v) / A(i\cdot m, i\cdot m)$.

Similarly, $bgt(i)(0)=b(i\cdot m)/A(i\cdot m, i\cdot m)$.

For $1\leq u\leq m-1$ we can compute $T(i,j)(u,v)$ and $bgt(i)(u)$ as follows:

\begin{equation}\label{eq5}
\begin{array}{l}
T(i,j)(j\neq i, u\geq 1, 0\leq v\leq m-1) = \\
-A(i\cdot m + u, j\cdot m + v) / A(i\cdot m + u, i\cdot m + u) + \\
\sum_{w=0}^{u-1}{(-A(i\cdot m + u, i\cdot m + w) / A(i\cdot m + u, i\cdot m + u)) \cdot T(i,j)(w,v)}
\end{array}
\end{equation}

\begin{equation}\label{eq6}
\begin{array}{l}
T(i,i)(u\geq 1,0\leq v\leq u) = \\
\sum_{w=0}^{u-1}{(-A(i\cdot m + u, i\cdot m + w) / A(i\cdot m + u, i\cdot m + u)) \cdot T(i,i)(w,v)}
\end{array}
\end{equation}

\begin{equation}\label{eq7}
\begin{array}{l}
T(i,i)(u\geq 1,u<v\leq m-1) = -A(i\cdot m + u, i\cdot m + v) / A(i\cdot m + u, i\cdot m + u) + \\
\sum_{w=0}^{u-1}{(-A(i\cdot m + u, i\cdot m + w) / A(i\cdot m + u, i\cdot m + u)) \cdot T(i,i)(w,v)}
\end{array}
\end{equation}

\begin{equation}\label{eq8}
\begin{array}{l}
bgt(i)(u\geq 1) = b(i\cdot m + u) / A(i\cdot m + u, i\cdot m + u) + \\
\sum_{w=0}^{u-1}{(-A(i\cdot m + u, i\cdot m + w) / A(i\cdot m + u, i\cdot m + u)) \cdot bgt(i)(w)}
\end{array}
\end{equation}

The computation of each value $T(i,j)(u,v)$ and $bgt(i)(u)$ can be easily parallelized to obtain a parallel runtime of $O(m/P+\log(P))$ for $P\leq m$ processors. Since all the values of a single row can be computed independently, each row $u$ of all the $T(i,j)$ matrices (for fixed $i$ and variable $j$) can be computed in $O(m\cdot n/P)$ time if $P\leq n$, or $O(m\cdot n/P+\log(P/n))$ time if $n<P\leq m\cdot n$. Since all the groups $i (0\leq i\leq n/m-1)$ are independent (there is no dependency between $T(i, j)$ and $T(i',j')$ with $i\neq i'$), we can compute one row from each block in parallel ($n/m$ rows in parallel), thus obtaining a parallel runtime of $O(n^2/P)$ if $P\leq n^2/m$, or $O(n^2/P+\log(P\cdot m/n^2))$ if $n^2/m<P\leq n^2$. In order to compute all the $m$ rows for each group we need to multiply the time complexities mentioned earlier by a factor of $m$.

A better approach when many processors $P$ are available can be derived if we consider the $(n/m)\cdot (n/m)$ matrices $A'(i,j)$ of size $m\cdot m$, where $A'(i,j)(u,v) = A(i\cdot m + u, j\cdot m + v)$ ($0\leq u, v\leq m-1$). If we let $D'(i)$, $-L'(i)$ and $-U'(i)$ be the diagonal part, strict lower part and strict upper part, respectively, of $A'(i,i)$, then $T(i,i)=(D'(i)-L'(i))^{-1}\cdot U'(i)$ (this is similar to Eq. \ref{eq2}). Similarly, $T(i,j)=(D'(i)-L'(i))^{-1}\cdot (-A'(i,j))$ (for $j\neq i$), and $bgt(i)=(D'(i)-L'(i))^{-1}\cdot bg(i)$. In order to compute all the matrices $T(i,j)$ (fixed $i$, $0\leq j\leq n/m-1$) we need to compute the inverse of a lower triangular matrix and then multiply the result (also a lower triangular matrix) by an upper triangular matrix ($U'(i)$) or a square matrix ($-A'(i,j)$). As mentioned earlier, there are sequential algorithms with $O(m^3)$ time complexities for both operations, but parallel algorithms also exist \cite{partrimatinv}, \cite{parmatinv}, \cite{parmatmult}. In the best case we can obtain an $O(\log^2(m))$ parallel runtime with $O(n^2\cdot m)$ processors available for computing all the blocks of $T$.

\subsection{The Implementation}
Once the matrix $T$ is computed, we can use the following algorithm for the Gauss-Seidel method. We will compute the values of the unknowns at the current iteration in increasing order of their groups. For each group we can achieve the maximum degree of parallelism possible, but the computation of the values for a given group may only be finalized after all the values from the previous groups were fully computed. At each step of our algorithm we will need to perform $O(n/m)$ independent matrix-vector multiplications (where each matrix has $m$ rows and $m$ columns and each vector has $m$ elements). Parallel algorithms for the matrix-vector multiplication problem were considered in \cite{parmatvec}. With $P\leq n$ processors available we can obtain a parallel runtime of $O(n\cdot m/P)$ (for all the matrix-vector multiplications). With $n<P\leq n\cdot m$ processors available we can obtain a parallel runtime of $O(n\cdot m/P+\log(P/n))$. Each iteration of our Gauss-Seidel algorithm consists of $n/m$ steps. Thus, the parallel runtime per iteration is $O(n^2/P)$ for $P\leq n$, or $O(n^2/P+(n/m)\cdot \log(P/n))$ for $n<P\leq n\cdot m$.

Besides the parallel matrix-vector multiplications there are several other simpler operations which need to be performed by the algorithm (e.g. initializing a vector in parallel, copying a vector in parallel, computing the difference of two vectors, computing the sum of the elements of a vector, etc.), which have parallel runtimes which are better than the parallel matrix-vector multiplication for the same number of available processors. Thus, the time required to perform matrix-vector multiplications dominates the running time of our algorithm.

An iteration of our parallel Gauss-Seidel method is presented in Algorithm \ref{pargsiter} while the full algorithm is presented in Algorithm \ref{pargs}. The algorithm maintains for each group $i$ two vectors of $m$ elements: $xgcurr(i)$ and $xgold(i)$. These vectors obey the following invariants. At the beginning of the step $i$ of the current iteration ($k+1$) we have: $xgcurr(i)$ already contains the values corresponding to $xg(k+1,i)$ from Eq. \ref{eq4}, and $xgold(i)$ contains the values of the unknowns from the previous iteration ($xg(k,i)$ from Eq. \ref{eq4}). At the end of the step $i$, $xgold(i)$ ends up containing the values of the unknowns at the end of the iteration ($xgcurr(i)$ is copied over into $xgold(i)$). Note that Algorithm \ref{pargsiter} follows a slightly different formulation than the one in Eq. \ref{eq4}, because when step $i$ starts, the values $xg(k+1,i)$ are already computed, while Eq. \ref{eq4} naturally implies that these values would need to be computed during the execution of the step.

\begin{algorithm}
\caption{Parallel Gauss-Seidel Iteration}
\label{pargsiter}
\begin{algorithmic}
\STATE $\varepsilon=0$
\FOR{$i=0$ to $n/m-1$}
\STATE Do in parallel $delta(i)=xgcurr(i)-xgold(i)$
\STATE Do in parallel $\varepsilon \mathrel{+}= \sum_{j=0..m-1}{|delta(i)(j)|}$
\STATE Do in parallel $xgold(i)=xgcurr(i)$
\FOR[Each iteration $j$ can be executed independently, in parallel.]{$j=0$ to $n/m-1$}
\STATE Compute in parallel $tmp(j,i) = T(j,i)\cdot delta(i)$
\STATE Do in parallel $xgcurr(j) \mathrel{+}= tmp(j,i)$
\ENDFOR
\ENDFOR
\end{algorithmic}
\end{algorithm}

\begin{algorithm}
\caption{Parallel Gauss-Seidel Algorithm}
\label{pargs}
\begin{algorithmic}
\STATE Compute in parallel the matrix $T$ and the vectors $bgt(i)$ ($0\leq i\leq n/m-1$)
\STATE Do in parallel $xgold(i)=0$ ($0\leq i\leq n/m-1$)
\STATE Do in parallel $xgcurr(i)=bgt(i)$ ($0\leq i\leq n/m-1$)
\STATE $\varepsilon = \infty$
\WHILE{$\varepsilon > \varepsilon_{convergence}$}
\STATE {\bf Parallel Gauss-Seidel Iteration}
\ENDWHILE
\end{algorithmic}
\end{algorithm}

\section{Conclusions}
\label{conclusions}

In this paper we presented a novel parallel Gauss-Seidel algorithm for solving dense systems of linear equations. The algorithm can obtain better parallel execution times than previous algorithms at the expense of more preprocessing time. The algorithm's preprocessing time can be tuned, so that the more preprocessing time we can accept, the more parallelism per iteration we can obtain.

Our algorithm may also be of particular interest when we need to solve multiple systems of linear equations which have the same matrix $A$ but different vectors $b$. Computing the inverses $(D'(i)-L'(i))^{-1}$ for each block $A'(i,i)$ and the matrix $T$ is the most time-consuming part of the preprocessing stage and only depends on the matrix $A$, so it only needs to be performed once. For each new vector $b$, we only need to compute the vectors $bgt(i)=(D'(i)-L'(i))^{-1}\cdot bg(i)$ via matrix-vector multiplications (or via Eq. \ref{eq8}).

\end{document}